# No vortex in straight flows — on the eigen-representations of velocity gradient $\nabla v$

Xiangyang Xu[1], Zhiwen Xu[1], Changxin Tang[1], Xiaohang Zhang[2], Wennan Zou[*1]

(1. Institute of Engineering Mechanics / Institute for Advanced Study / Institute of Photovoltaics, Nanchang University, Nanchang 330031, China; 2. School of Science, Nanchang Institute of Technology, Nanchang 330099, China)

**ABSTRACT**: Velocity gradient is the basis of many vortex recognition methods, such as Q criterion, Δ criterion, $\lambda_2$ criterion, $\lambda_{ci}$ criterion and Ω criterion, etc.. Except the $\lambda_{ci}$ criterion, all these criterions recognize vortices by designing various invariants, based on the Helmholtz decomposition that decomposes velocity gradient into strain rate and spin. In recent years, the intuition of "no vortex in straight flows" has promoted people to analyze the vortex state directly from the velocity gradient, in which vortex can be distinguished from the situation that the velocity gradient has couple complex eigenvalues. A specious viewpoint to adopt the simple shear as an independent flow mode was emphasized by many authors, among them, Kolář (2004) proposed the triple decomposition of motion by extracting a so-called 'effective' pure shearing motion; Li *et al*. (2014) introduced the so-called quaternion decomposition of velocity gradient and proposed the concept of eigen rotation; Liu *et al*. (2016) further mined the characteristic information of velocity gradient and put forward an effective algorithm of Liutex (namely eigen rotation), and then developed the vortex recognition method. However, there is another explanation for the increasingly clear representation of velocity gradient, that is the local streamline pattern based on critical-point theory. In this paper, the tensorial expressions of the right/left real Schur forms of velocity gradient are clarified from the characteristic problem of $\nabla v$. The relations between the involved parameters are derived and numerically verified. Comparing with the geometrical features of local streamline pattern, we confirm that the parameters in the right eigen-representation based on the right real Schur form of velocity gradient have good meanings to reveal the local streamline pattern. Some illustrative examples from the DNS data are presented.

## 1 Introduction

The intuition of no vortex in straight flows becomes strong in the understanding of vortex identification. Lugt (1983) presented that a vortex is the rotating motion of a multitude of material particles around a common center, while Robinson (1991) proposed that a vortex exists when instantaneous streamlines mapped onto a plane normal to the vortex core exhibit a roughly circular or spiral pattern, when viewed from a reference frame moving with the center of the vortex core. Therefore, the fact - vorticity cannot distinguish between pure shearing motions and the actual swirling motion of a vortex (Jeong and Hussain, 1995; Kida and Miura, 1998; Cucitore *et al.*, 1999) – has won support among the people to develop other ways instead of vorticity to identify a vortex. In the framework of classical flow theory, the vorticity, indicating an average angular velocity of fluid elements, appears as one of the unique natural choices for a vortex-identification criterial measure.

The fatal flaw of using vorticity to indicate the vortex is that we have to admit the vortex in straight flows. She *et al*. (1990) found the tube-like feature of strong vortices in the DNS turbulence, Saffman (1992) added that "we shall use this term to denote any finite volume of vorticity immersed in irrotational fluid", and recently Wu and Yang (2020) proposed that "vortex is a specific region of vorticity field with tubular structure", and tried to consider the dynamic mechanism within the definition of vortex. In order to consider the effect of strain rate in addition to vorticity, several methods were developed to analyze the vortex under the requirement of the Galilean invariance (Jiang *et al*., 2005; Epps, 2017). Most of them are derived from the velocity gradient $d = \nabla v$, where the methods based on the scalar invariants (eigenvalues) of velocity gradient (or its sum decomposition) include Q criterion (Okubo, 1970; Hunt *et al*., 1988; Weiss, 1991), Δ criterion (Chong and Perry, 1990), $\lambda_2$

---

criterion (Jeong and Hussain, 1995) and $\lambda_{ci}$ criterion (Zhou et al., 1999; Chakraborty et al., 2005), among them only the $\lambda_{ci}$ criterion has nothing to do with the Helmholtz decomposition.

After a lot of theoretical and practical exploration, the application of complex measures derived from $d$ has already revealed its importance in the analysis of vortical structures in complicated flows. But the flow mechanism other than the Helmholtz decomposition seems to be difficult to construct. Kolář (2004, 2007) proposed a triple decomposition by extracting of a so-called 'effective' pure shearing motion, limited to planar flows. Li et al. (2014) presented the quadruple decomposition of velocity gradient, namely dilatation, axial deformation along the principal axes of the strain-range sensor, planar motion, and pure shearing. Liu and his coworkers (2018, 2019) realized such a decomposition is actually based on the real Schur form (Golub and van Loan, 2013) of the velocity gradient, and constructed a systemic criterion called Liutex/Rortex. But the abandon of the Helmholtz decomposition means the mechanism of viscous interaction coming from the strain rate must be modified.

An alternative idea making use of velocity gradient to catch flow patterns stems from critical point theory (Dallmann, 1983; Vollmers et al., 1983; Perry and Chong, 1987). Since Perry and Chong (1987) pointed out the usefulness of critical-point concepts in the understanding of flow patterns, Chong et al. (1990) proposed the use of the region where a couple of complex eigenvalue implies the appearance of a vortex, Zhou et al. (1999) used the imaginary part of the complex eigenvalue of velocity gradient, and presented the local streamline pattern (LSP) to visualize a vortex, Wang et al. (2019) also investigated the imaginary part $\lambda_{ci}$ of the complex eigenvalue of the velocity gradient as the pseudo-time average angular velocity of a trajectory moving circularly or spirally around the axis. In practice, many people thought it necessary to combine this methodology with the concentration of vorticity magnitude, namely dividing the vorticity into a rotation part and a shear part, to obtain a reasonable shape for the vortex.

In this paper, we will focus on the vortex identification methods derived from the real Schur form of velocity gradient. In section 2, three studies starting from the complex eigenvalues of velocity gradient are summarized and compared with each other. In section 3, we focus on the tensorial representations and the relationship between different representations. The LSPs are classified in section 4, and the correspondence between the geometrical features and the parameters in the eigen-representations is investigated. Discussion and Case study are presented in section 5, while a brief conclusion is given in section 6.

## 2. Vortex recognition methods from velocity gradient with complex eigenvalues

For simplicity, the fluid is uniform and incompressible in this paper, that means, the divergence of velocity vanishes everywhere. For a three-dimensional flow, the matrix form of velocity gradient is written as

$$\nabla \boldsymbol{v} = \begin{bmatrix} \frac{\partial u}{\partial x} & \frac{\partial v}{\partial x} & \frac{\partial w}{\partial x} \\ \frac{\partial u}{\partial y} & \frac{\partial v}{\partial y} & \frac{\partial w}{\partial y} \\ \frac{\partial u}{\partial z} & \frac{\partial v}{\partial z} & \frac{\partial w}{\partial z} \end{bmatrix} = \begin{bmatrix} d_{11} & d_{12} & d_{13} \\ d_{21} & d_{22} & d_{23} \\ d_{31} & d_{32} & d_{33} \end{bmatrix} \equiv \boldsymbol{d}. \tag{1}$$

When the velocity gradient has only one real eigenvalue, Zhou et al. (1999) wrote the velocity gradient by[2]

$$\boldsymbol{d}^T = \begin{bmatrix} \boldsymbol{v}_{cr} & \boldsymbol{v}_{ci} & \boldsymbol{v}_r \end{bmatrix} \begin{bmatrix} -\frac{1}{2}\lambda_r & \lambda_i & 0 \\ -\lambda_i & -\frac{1}{2}\lambda_r & 0 \\ 0 & 0 & \lambda_r \end{bmatrix} \begin{bmatrix} \boldsymbol{v}_{cr} & \boldsymbol{v}_{ci} & \boldsymbol{v}_r \end{bmatrix}^{-1}, \tag{2}$$

where $\lambda_r$ is the real eigenvalue with $\boldsymbol{v}_r$ as its eigenvector, while the two conjugate complex eigenvalues $-\frac{1}{2}\lambda_r \pm i\lambda_i$ have corresponding eigenvectors $\boldsymbol{v}_{cr} \pm i\boldsymbol{v}_{ci}$. In the local affine coordinate system $\{y_1, y_2, y_3\}$ defined by bases $\{\boldsymbol{v}_{cr}, \boldsymbol{v}_{ci}, \boldsymbol{v}_r\}$, the local streamline starting at point $(x_1^0, x_2^0, x_3^0)$ can be solved from the velocity field $\boldsymbol{v} = \frac{d\mathbf{r}}{dt} = \boldsymbol{d}^T \cdot \mathbf{r}$ as

$$y_3(t) = x_3^0 e^{\lambda_r t}, \tag{3.1}$$

$$y_1(t) = e^{-\frac{1}{2}\lambda_r t}\left[x_1^0 \cos(\lambda_i t) + x_2^0 \sin(\lambda_i t)\right], \tag{3.2}$$

$$y_2(t) = e^{-\frac{1}{2}\lambda_r t}\left[x_2^0 \cos(\lambda_i t) - x_1^0 \sin(\lambda_i t)\right]. \tag{3.3}$$

Zhou et al. (1999) pointed out that "the local flow is either stretched or compressed along the axis $\boldsymbol{v}_r$, while on the plane spanned by the vectors $v_{cr}$ and $v_{ci}$, the flow is swirling", as shown in Fig. 1. That means the normal of plane $(v_{cr}, v_{ci})$, or the right eigenvector of the real eigenvalue of $d$, is referred to as the rotation direction of local vortex, and the left eigenvector $v_r$ indicates the direction we will define as the extension direction of local vortex. In addition, Zhou et al. (1999) used the imaginary part of the complex eigenvalue pair as the local swirling strength of the vortex.

---
[2] According to the matrix notation, say the equation (1) in Chong et al. (1990), the velocity gradient used by Zhou et al. (1999), is actually $\boldsymbol{v}\nabla$, so we denote it by $\boldsymbol{d}^T$ in this paper. We also rearrange the order of eigenvalues and eigenvectors.

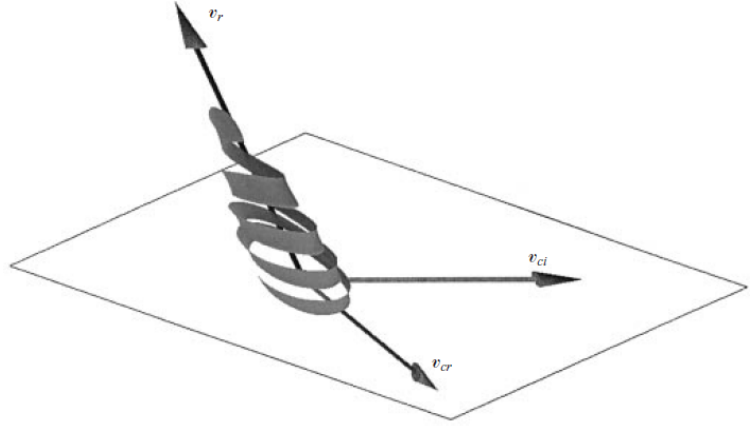

**Fig. 1.** The LSP of velocity gradient tensor pointed out by Zhou *et al* (1999).

Unlike the two-dimensional flow (Kolář, 2004, 2007), in which the rotation and extension directions of the vortex are the same and definitely perpendicular to the plane, in the three-dimensional flow both rotation and extension directions of the vortex are to be determined. According to the Schur theorem (Golub and von Loan, 2013), there are real Schur forms for a real matrix with the real eigenvalue $\lambda_r$ (the only one if not specified), say

$$d = P^{*T} D_R P^* = P^T D_L P \tag{4.1}$$

with

$$D_L = \begin{bmatrix} D_{11} & D_{12} & D_{13} \\ D_{21} & D_{22} & D_{23} \\ 0 & 0 & D_{33} \end{bmatrix}, D_R = \begin{bmatrix} D_{11}^* & D_{12}^* & 0 \\ D_{21}^* & D_{22}^* & 0 \\ D_{31}^* & D_{32}^* & D_{33}^* \end{bmatrix}, \tag{4.2}$$

and $D_{33}^* = D_{33} = \lambda_r$, $n_3 \cdot d = \lambda_r n_3$, $d \cdot m_3 = \lambda_r m_3$. It is easy to testify that $n_3$ is equivalent to $v_r$ while $m_3$ indicates the axis of $(v_{cr}, v_{ci})$.

After Kolář's proposition and practice for two-dimensional flows, Li et al. (2014) first introduced an orthonormal frame $\{n_1, n_2, n_3\}$ (right hand if not specified) to express $D_L$ as in Table 1, and called $\psi$ the proper rotation. Years later, based on the same left real Schur form (see Table 1), Liu et al. (2018) proposed the rigid-body rotation vector by combining the parameter $\phi$ with the direction $n_3$, and presented an effective algorithm for calculating the involved parameters and achieved a lot of applications through cooperation.

As listed in Table 1, Zhou, Li and Liu started from the same (left) characteristic problem of velocity gradient tensor, Li and Liu introduced the same orthonormal frame, used the same parameter to characterize the strength of rotation, but both of them paid no attention to the extension direction.

**Table 1.** Three representations of velocity gradient tensor $\nabla v$ with one real eigenvalue $\lambda_r$ and its left eigenvector $v_r$ or $n_3$

| | ① Zhou *et al.* ($\lambda_{ci}$) | ② Li *et al.* (Proper rotation) | ③ Liu *et al.* (Liutex) | Remarks |
|---|---|---|---|---|
| Representation | $\begin{bmatrix} -\frac{1}{2}\lambda_r & \lambda_i & 0 \\ -\lambda_i & -\frac{1}{2}\lambda_r & 0 \\ 0 & 0 & \lambda_r \end{bmatrix}$ | $\begin{bmatrix} -\frac{1}{2}\lambda_r & \psi+\gamma & \beta \\ -\psi & -\frac{1}{2}\lambda_r & \alpha \\ 0 & 0 & \lambda_r \end{bmatrix}$ | $\begin{bmatrix} -\frac{1}{2}\lambda_r & \phi+s & \xi \\ -\phi & -\frac{1}{2}\lambda_r & \eta \\ 0 & 0 & \lambda_r \end{bmatrix}$ | ② and ③ are orthonormal frames, ① is affine frame |
| Rotation axis | the axis of plane $(v_{cr}, v_{ci})$ | $n_3$ | $n_3$ | ① is rotation plane, ② and ③ are the same |
| Rotation strength | $\lambda_i$ | $\psi$ | $\phi$ | ② and ③ are the same |
| Extension direction | Stretch/compress direction $n_3$ | – | – | No definition in ② and ③ |

## 3. Three forms of tensorial representation for the velocity gradient

Dividing the fluid domain into the regions with or without vortex is the primary objective in vortex identification, which can be worked out by the feature that the characteristic polynomial of velocity gradient tensor has complex roots or not.

### 3.1 Spectral representation under the affine frame in the vortex region

For the velocity gradient $\boldsymbol{d}$, the characteristic equations of its left characteristic problem $\boldsymbol{N} \cdot \boldsymbol{d} = \lambda \boldsymbol{N}$ and its right characteristic problem $\boldsymbol{d} \cdot \boldsymbol{N} = \lambda \boldsymbol{N}$ are the same. Assume that the characteristic polynomial has eigenvalues $\left\{-\frac{1}{2}\lambda_3 \pm \iota \beta, \lambda_3\right\}$ with the unit imaginary number $\iota = \sqrt{-1}$, and the corresponding right eigenvectors are $\{\boldsymbol{N}_1 \pm \iota \boldsymbol{N}_2, \boldsymbol{N}_3\}$ satisfying

$$\boldsymbol{d} \cdot \boldsymbol{N}_3 = \lambda_3 \boldsymbol{N}_3, \qquad \boldsymbol{d} \cdot (\boldsymbol{N}_1 \pm \iota \boldsymbol{N}_2) = \left(-\frac{1}{2}\lambda_3 \pm \iota \beta\right)(\boldsymbol{N}_1 \pm \iota \boldsymbol{N}_2); \tag{5}$$

and the corresponding left eigenvectors are $\{\boldsymbol{N}^1 \mp \iota \boldsymbol{N}^2, \boldsymbol{N}^3\}$ satisfying

$$\boldsymbol{N}^3 \cdot \boldsymbol{d} = \lambda_3 \boldsymbol{N}^3, \qquad (\boldsymbol{N}^1 \mp \iota \boldsymbol{N}^2) \cdot \boldsymbol{d} = \left(-\frac{1}{2}\lambda_3 \pm \iota \beta\right)(\boldsymbol{N}^1 \mp \iota \boldsymbol{N}^2). \tag{6}$$

From (5) and (6), the spectral representation of velocity gradient tensor

$$\boldsymbol{d} = \lambda_3 \boldsymbol{N}_3 \otimes \boldsymbol{N}^3 + \mathrm{Re}\left[\left(-\frac{1}{2}\lambda_3 + \iota \beta\right)(\boldsymbol{N}_1 + \iota \boldsymbol{N}_2) \otimes (\boldsymbol{N}^1 - \iota \boldsymbol{N}^2)\right], \tag{7.1}$$

with the combination of two right-handed frames $\{\boldsymbol{N}_1, \boldsymbol{N}_2, \boldsymbol{N}_3\}$ and $\{\boldsymbol{N}^1, \boldsymbol{N}^2, \boldsymbol{N}^3\}$ is form-invariant by requiring the dual relations $\boldsymbol{N}^i \cdot \boldsymbol{N}_j = \delta^i_j$ [3]. The dual relations show that the two right-handed frames $\{\boldsymbol{N}_1, \boldsymbol{N}_2, \boldsymbol{N}_3\}$ and $\{\boldsymbol{N}^1, \boldsymbol{N}^2, \boldsymbol{N}^3\}$ are uniquely interdependent in the spectral representation. An equivalent expression of (7.1) is

$$\boldsymbol{d} = \lambda_3 \boldsymbol{N}_3 \otimes \boldsymbol{N}^3 - \frac{1}{2}\lambda_3(\boldsymbol{N}_1 \otimes \boldsymbol{N}^1 + \boldsymbol{N}_2 \otimes \boldsymbol{N}^2) + \beta(\boldsymbol{N}_1 \otimes \boldsymbol{N}^2 - \boldsymbol{N}_2 \otimes \boldsymbol{N}^1), \tag{7.2}$$

which coincides with the matrix expression given by Zhou *et al* (1999).

It is obvious in the spectral representation (7) that the real eigenvector, $\boldsymbol{N}_3$ or $\boldsymbol{N}^3$, can be determined to be only one non-zero real factor difference as long as the tensor product $\boldsymbol{N}_3 \otimes \boldsymbol{N}^3$ is constant, while the complex eigenvector, $\boldsymbol{N}_1 + \iota \boldsymbol{N}_2$ or $\boldsymbol{N}^1 + \iota \boldsymbol{N}^2$, can be determined to be only one non-zero complex factor difference as long as the tensor product $(\boldsymbol{N}_1 + \iota \boldsymbol{N}_2) \otimes (\boldsymbol{N}^1 - \iota \boldsymbol{N}^2)$ is invariant. Let alone the size changes of $\boldsymbol{N}_1, \boldsymbol{N}_2, \boldsymbol{N}_3$, consider the basic transforms including the sign changes of $\boldsymbol{N}_1, \boldsymbol{N}_2, \boldsymbol{N}_3$ and exchanges of $\boldsymbol{N}_1$ and $\boldsymbol{N}_2$, there are 16 possible frames where half of them are right-handed. Among the 8 right-handed frames, four of them $\{\boldsymbol{N}_1, \boldsymbol{N}_2, \boldsymbol{N}_3\}$, $\{-\boldsymbol{N}_1, -\boldsymbol{N}_2, \boldsymbol{N}_3\}$, $\{\boldsymbol{N}_2, -\boldsymbol{N}_1, \boldsymbol{N}_3\}$ and $\{-\boldsymbol{N}_2, \boldsymbol{N}_1, \boldsymbol{N}_3\}$ result in the same spectral representation as (7), and if $\boldsymbol{N}_3$ changes to its opposite $-\boldsymbol{N}_3$, the remaining four frames $\{\boldsymbol{N}_1, -\boldsymbol{N}_2, -\boldsymbol{N}_3\}$, $\{-\boldsymbol{N}_1, \boldsymbol{N}_2, -\boldsymbol{N}_3\}$, $\{\boldsymbol{N}_2, \boldsymbol{N}_1, -\boldsymbol{N}_3\}$ and $\{-\boldsymbol{N}_2, -\boldsymbol{N}_1, -\boldsymbol{N}_3\}$ will yield the same form of representation but $\beta$ changes to $-\beta$. Therefore, if we limit $\beta$ to a positive number, the directions of real eigenvectors $\boldsymbol{N}_3$ and $\boldsymbol{N}^3$ are uniquely determined, and there are four equivalent right-handed frames for the spectral representation (7).

### 3.2 Eigen-representations under the orthonormal frames in the vortex region

In this subsection, we will denote the orthonormal frame based on $\boldsymbol{N}^3$ by the left orthonormal frame, and refer to the tensorial form of the matrix expression of velocity gradient under this frame (Li *et al.*, 2014; Liu *et al.*, 2019) as the left eigen-representation. We propose the right eigen-representation under the right orthonormal frame based on $\boldsymbol{N}_3$, and derive the relation between it and the spectral representation (7).

**Theorem 1**: For the case that the characteristic polynomial of velocity gradient tensor has complex roots, there exists a *unique* right-handed orthonormal frame $\{\boldsymbol{m}_1, \boldsymbol{m}_2, \boldsymbol{m}_3\}$, and the right eigen-representation

$$\boldsymbol{d} = \lambda_3 \lfloor \boldsymbol{m}_3 \otimes \boldsymbol{m}_3 \rfloor + (R + \tau_3)\boldsymbol{m}_1 \otimes \boldsymbol{m}_2 - R\boldsymbol{m}_2 \otimes \boldsymbol{m}_1 + \boldsymbol{m}_3 \otimes (\tau_1 \boldsymbol{m}_1 + \tau_2 \boldsymbol{m}_2) \tag{8}$$

under $\{\boldsymbol{m}_1, \boldsymbol{m}_2, \boldsymbol{m}_3\}$, with $R > 0$, $\tau_3 > 0$ and $\tau_1 > 0$ (or $\tau_2 > 0$ when $\tau_1 = 0$), where the symmetric traceless base $\lfloor \boldsymbol{m}_3 \otimes \boldsymbol{m}_3 \rfloor$ is defined by $\lfloor \boldsymbol{m}_3 \otimes \boldsymbol{m}_3 \rfloor \equiv \boldsymbol{m}_3 \otimes \boldsymbol{m}_3 - \frac{1}{2}\boldsymbol{m}_1 \otimes \boldsymbol{m}_1 - \frac{1}{2}\boldsymbol{m}_2 \otimes \boldsymbol{m}_2$. If the coefficient of $\boldsymbol{m}_2 \otimes \boldsymbol{m}_1$ is negative, $\boldsymbol{m}_3$ cannot change to $-\boldsymbol{m}_3$; If $\tau_1 = \tau_2 = 0$, $\{-\boldsymbol{m}_1, -\boldsymbol{m}_2, \boldsymbol{m}_3\}$ is an equivalent frame; if $\tau_3 = 0$, $\{-\boldsymbol{m}_2, \boldsymbol{m}_1, \boldsymbol{m}_3\}$ also becomes an equivalent frame.

---

[3] The dual conditions can be solved explicitly. Assume that $\{\boldsymbol{N}_1, \boldsymbol{N}_2, \boldsymbol{N}_3\}$ with the triple product $[\boldsymbol{N}_1, \boldsymbol{N}_2, \boldsymbol{N}_3] = (\boldsymbol{N}_1 \times \boldsymbol{N}_2) \cdot \boldsymbol{N}_3 = \sqrt{g} > 0$ is known, the corresponding $\{\boldsymbol{N}^1, \boldsymbol{N}^2, \boldsymbol{N}^3\}$ can be expressed by $\boldsymbol{N}^i = \frac{1}{2}\sqrt{g}^{-1} \epsilon^{ijk} \boldsymbol{N}_j \times \boldsymbol{N}_k$, and vice versa.

In order to guarantee the uniqueness of correspondence between (7) and (8), we need to regulate the eigenvectors in (7). As mentioned, the four right-handed frames $\{N_1, N_2, N_3\}$, $\{-N_1, -N_2, N_3\}$, $\{N_2, -N_1, N_3\}$ and $\{-N_2, N_1, N_3\}$ are indistinguishable according to the basic eigen-parameters $\lambda_3$ and $\beta$. The requirement of an obtuse angle between $N^1$ and $N^2$ to distinguish $\{N_1, N_2, N_3\}$ and $\{N_2, -N_1, N_3\}$, and the sign of $m_1 \cdot N_1$ or $m_3 \cdot d \cdot m_1$ to distinguish $\{N_1, N_2, N_3\}$ and $\{-N_1, -N_2, N_3\}$. Because of the dual relations, the treatment to the right-handed frames $\{N_1, N_2, N_3\}$ is also workable to the right-handed frames $\{N^1, N^2, N^3\}$.

Now we start from (8) to achieve the spectral representation (7). Since the bases $\{m_1, m_2, m_3\}$ and the parameters $\lambda_3, R, \tau_1, \tau_2, \tau_3$ are known, we first have $N_3 = m_3$ and $\beta = \sqrt{R(R+\tau_3)}$. To make $N^1$ and $N^2$ unique, we regulate them to have the same size, and the included angle between $N^1$ and $m_1$ to be acute. Because both $N^1$ and $N^2$ are in the plane normal to $N_3$, they can be expressed by unit vectors $m_1$ and $m_2$, while their sizes are further required to meet the requirement $N^1 \times N^2 = m_3$. From (8) and the characteristic relation $(N^1 - \iota N^2) \cdot d = \left(-\frac{1}{2}\lambda_3 + \iota\beta\right)(N^1 - \iota N^2)$, we obtain

$$N^1 = \sqrt{\frac{R\beta}{2}}\left(\frac{m_1}{\beta} - \frac{m_2}{R}\right), N^2 = \sqrt{\frac{R\beta}{2}}\left(\frac{m_1}{\beta} + \frac{m_2}{R}\right), \quad (9.1)$$

or

$$N^1 - \iota N^2 = (1-\iota)\sqrt{\frac{R\beta}{2}}\left(\frac{m_1}{\beta} - \iota\frac{m_2}{R}\right). \quad (9.2)$$

Assume that $N^3 = m_3 + C_1 m_1 + C_2 m_2$, combining with $N^3 \cdot d = \lambda_3 N^3$ and (8) results in $\begin{pmatrix} \frac{3}{2}\lambda_3 & R \\ -\frac{\beta^2}{R} & \frac{3}{2}\lambda_3 \end{pmatrix}\begin{pmatrix} C_1 \\ C_2 \end{pmatrix} = \begin{pmatrix} \tau_1 \\ \tau_2 \end{pmatrix}$, which can be solved to get

$$N^3 = m_3 + \frac{\frac{3}{2}\lambda_3 \tau_1 - R\tau_2}{\frac{9}{4}\lambda_3^2 + \beta^2} m_1 + \frac{\frac{3}{2}\lambda_3 \tau_2 + \frac{\beta^2}{R}\tau_1}{\frac{9}{4}\lambda_3^2 + \beta^2} m_2. \quad (10)$$

Finally, making use of the orthogonality $N_1 \cdot N^1 = N_2 \cdot N^2 = 1$ and $N_1 \cdot N^2 = N_2 \cdot N^1 = (N_1 + \iota N_2) \cdot N^3 = 0$, we have

$$N_1 + \iota N_2 = (1+\iota)\sqrt{\frac{1}{2R\beta}}\left(\beta m_1 + \iota R m_2 - \frac{\beta\tau_1 + \iota R\tau_2}{\frac{3}{2}\lambda_3 - \iota\beta} m_3\right). \quad (11)$$

The above results for eigenvectors can be testified by substituting them into the spectral representation (7), which will yield the right eigen-representation (8) of velocity gradient under the orthonormal frame $\{m_1, m_2, m_3\}$.

The inverse relations can be obtained as follows. When the eigenvalues $\left\{-\frac{1}{2}\lambda_3 \pm \iota\beta, \lambda_3\right\}$ are given, the right eigenvectors $\{N^1 \pm \iota N^2, N^3\}$ are determined to some extent. We further choose the right-handed frame $\{N^1, N^2, N^3\}$ by making the coefficient of $N_1 \otimes N^2$ in (7.2) positive (If not so, use the equivalent frame $\{N^1, -N^2, -N^3\}$ instead) and an obtuse angle between $N^1$ and $N^2$ (If not so, use the equivalent frame $\{-N^2, N^1, N^3\}$ instead); then normalize $N_3$ to a unit vector and isomorphize $N^1$ and $N^2$ in advance through a rotation of $\theta$ around the axis $N_3$ so that the real part and the imaginary part of $(N^1 + \iota N^2)e^{-\iota\theta}$ have the same size. Now, if all these requirements are met, namely (a) $\{N^1, N^2, N^3\}$ is right-handed, (b) $N_3$ is a unit vector, (c) $N^1$ and $N^2$ are isomorphized and their included angle is obtuse, it is easy to calculate the orthonormal frame $\{m_1, m_2, m_3\}$ and the parameters $\lambda_3, \beta, R, \tau_1, \tau_2, \tau_3$ in the right eigen-decomposition (8):

1) Set $m_3$ being the normalized $N_3$;
2) Calculate $R$ and $m_{1,2}$ from the isomorphized $N^1, N^2$ by $R = \beta\frac{|N^1+N^2|}{|N^1-N^2|} \leq \beta$, $m_{1,2} = \frac{N^2 \pm N^1}{|N^1 \pm N^2|}$;
3) Calculate other parameters as $\tau_3 = (\beta^2 - R^2)/R$, $\tau_1 = m_3 \cdot d \cdot m_1, \tau_2 = m_3 \cdot d \cdot m_2$;
4) Finally, if $\tau_1 < 0$, or $\tau_1 = 0, \tau_2 < 0$, using the equivalent frame $\{-N_1, -N_2, N_3\}$ to recalculate $m_{1,2}$, and so on.

Similarly, in order to build up the relation between the left eigen-representation under the right-handed orthonormal frame $\{n_1, n_2, n_3\}$

$$d = \lambda_3[n_3 \otimes n_3] + (R^* + \tau_3^*)n_1 \otimes n_2 - R^* n_2 \otimes n_1 + (\tau_1^* n_1 + \tau_2^* n_2) \otimes n_3 \quad (12)$$

and the spectral representation (7), we set $N^3$ to be $n_3$, and $N_1$ and $N_2$

$$N_1 = \sqrt{\frac{R^*\beta}{2}}\left(\frac{n_1}{R^*} - \frac{n_2}{\beta}\right), N_2 = \sqrt{\frac{R^*\beta}{2}}\left(\frac{n_1}{R^*} + \frac{n_2}{\beta}\right), \quad (13)$$

or

$$N_1 + \iota N_2 = (1 + \iota)\sqrt{\frac{R^*\beta}{2}}\left(\frac{n_1}{R^*} + \iota\frac{n_2}{\beta}\right) \quad (14)$$

satisfying the conditions: $(N_1, N_2) \perp N^3$, $N_1 \times N_2 = n_3$, $\boldsymbol{d} \cdot (N_1 + \iota N_2) = \left(-\frac{1}{2}\lambda_3 + \iota\beta\right)(N_1 + \iota N_2)$, and the included angle between $N_1$ and $n_1$ being acute. Assume that $N_3 = n_3 + D_1 n_1 + D_2 n_2$, the linear equations $\begin{pmatrix} \frac{3}{2}\lambda_3 & -\frac{\beta^2}{R^*} \\ R^* & \frac{3}{2}\lambda_3 \end{pmatrix}\begin{pmatrix} D_1 \\ D_2 \end{pmatrix} = \begin{pmatrix} \tau_1^* \\ \tau_2^* \end{pmatrix}$ from $\boldsymbol{d} \cdot N_3 = \lambda_3 N_3$ can be solved as

$$N_3 = n_3 + \frac{\frac{3}{2}\lambda_3\tau_1^* + \frac{\beta^2}{R^*}\tau_2^*}{\frac{9}{4}\lambda_3^2 + \beta^2}n_1 + \frac{\frac{3}{2}\lambda_3\tau_2^* - R^*\tau_1^*}{\frac{9}{4}\lambda_3^2 + \beta^2}n_2. \quad (15)$$

Due to the orthogonality with $N_1 + \iota N_2$, we set $N^1 - \iota N^2 = (1-\iota)\sqrt{\frac{1}{2R^*\beta}}(R^* n_1 - \iota\beta n_2 + D n_3)$ with an undetermined parameter $D$, and figure out it from the relation $(N^1 - \iota N^2) \cdot N_3 = 0$, that is

$$N^1 - \iota N^2 = (1-\iota)\sqrt{\frac{1}{2R^*\beta}}\left(R^* n_1 - \iota\beta n_2 - \frac{R^*\tau_1^* - \iota\beta\tau_2^*}{\frac{3}{2}\lambda_3 - \iota\beta}n_3\right). \quad (16)$$

The expansion after the substitution of the above results into the spectral representation (7) also results in the left eigen-representation (12) of velocity gradient under the orthonormal frame $\{n_1, n_2, n_3\}$. The inverse relations of left eigen-representation can also be obtained in a similar way to those of right eigen-representation.

Both right and left eigen-representation are unique in the case of the characteristic equation having complex roots. Since these two decompositions share the same eigenvalues, it is easy to find that the parameters $\lambda_3, \beta$ in their representations must be completely identical, but other parameters and the orthonormal frame are in general not the same. Therefore, we denote parameters in the left eigen-representation by $R^*, \tau_1^*, \tau_2^*, \tau_3^*$ instead of $R, \tau_1, \tau_2, \tau_3$. It can be deduced that $R$ and $R^*$ have the following transformation relations:

$$R^* = R\sqrt{\left(1 + \frac{\tau_3}{R}\right)\frac{c_0^2(1+c_1^2) + c_0^{-2}(1+c_2^2) - \sqrt{4c_1^2c_2^2 + [c_0^2(1+c_1^2) - c_0^{-2}(1+c_2^2)]^2}}{c_0^2(1+c_1^2) + c_0^{-2}(1+c_2^2) + \sqrt{4c_1^2c_2^2 + [c_0^2(1+c_1^2) - c_0^{-2}(1+c_2^2)]^2}}}, \quad (17.1)$$

$$c_0 = \sqrt{\frac{\beta}{R}}, \quad c_1 = \frac{\frac{3}{2}\lambda_3\tau_1 - R\tau_2}{\frac{9}{4}\lambda_3^2 + \beta^2}, \quad c_2 = \frac{\frac{3}{2}\lambda_3\tau_2 + \frac{\beta^2}{R}\tau_1}{\frac{9}{4}\lambda_3^2 + \beta^2}; \quad (17.2)$$

$$R = R^*\sqrt{\left(1 + \frac{\tau_3^*}{R^*}\right)\frac{c_0^{*2}(1+c_1^{*2}) + c_0^{*-2}(1+c_2^{*2}) - \sqrt{4c_1^{*2}c_2^{*2} + [c_0^{*2}(1+c_1^{*2}) - c_0^{*-2}(1+c_2^{*2})]^2}}{c_0^{*2}(1+c_1^{*2}) + c_0^{*-2}(1+c_2^{*2}) + \sqrt{4c_1^{*2}c_2^{*2} + [c_0^{*2}(1+c_1^{*2}) - c_0^{*-2}(1+c_2^{*2})]^2}}}, \quad (18.1)$$

$$c_0^* = \sqrt{\frac{\beta}{R^*}}, \quad c_1^* = \frac{\frac{3}{2}\lambda_3\tau_2^* - R^*\tau_1^*}{\frac{9}{4}\lambda_3^2 + \beta^2}, \quad c_2^* = \frac{\frac{3}{2}\lambda_3\tau_1^* + \frac{\beta^2}{R^*}\tau_2^*}{\frac{9}{4}\lambda_3^2 + \beta^2}. \quad (18.2)$$

The deduction process and more relations are given in the Appendix A.

### 3.3 Eigen-representation under the right orthonormal frame in the non-vortex region

The special case of the characteristic equation having three real roots $\{\lambda_1, \lambda_2, \lambda_3\}$ $(\lambda_1 + \lambda_2 + \lambda_3 = 0)$ is rarely discussed in the literature. We find its tensorial representation can be written in a similar form under an orthogonal frame. For simplicity, consider three roots be distinct and $\lambda_3$ be the one with the largest module. Assume

$$\boldsymbol{d} = \lambda_1 N_1 \otimes N^1 + \lambda_2 N_2 \otimes N^2 + \lambda_3 N_3 \otimes N^3 \quad (19)$$

being the spectral representation of this case, where both $\{N^1, N^2, N^3\}$ and $\{N_1, N_2, N_3\}$ are right-handed and there is an obtuse included angle between $N^1$ and $N^2$. Further normalize $N_3$ of $\lambda_3$ to be unit base $m_3$, make use of $\lambda_{1,2} = -\frac{1}{2}\lambda_3 \pm \beta$, define $\beta^2 = R(R + \tau_3)$, and isomorphize $N^1$ and $N^2$ as

$$N^1 = \sqrt{\frac{R\beta}{2}}\left(\frac{m_1}{\beta} - \frac{m_2}{R}\right), \qquad N^2 = \sqrt{\frac{R\beta}{2}}\left(\frac{m_1}{\beta} + \frac{m_2}{R}\right), \tag{20}$$

we can get the orthonormal frame $\{m_1, m_2, m_3\}$, and express the other bases satisfying the orthogonal relations $N^i \cdot N_j = \delta^i_j$ as

$$N_1 = \sqrt{\frac{1}{2R\beta}}\left(\beta m_1 - R m_2 - \frac{\beta\tau_1 - R\tau_2}{\frac{3}{2}\lambda_3 - \beta} m_3\right), N_2 = \sqrt{\frac{1}{2R\beta}}\left(\beta m_1 + R m_2 - \frac{\beta\tau_1 + R\tau_2}{\frac{3}{2}\lambda_3 + \beta} m_3\right) \tag{21}$$

$$N^3 = m_3 + \frac{\frac{3}{2}\lambda_3\tau_1 - R\tau_2}{\frac{9}{4}\lambda_3^2 - \beta^2} m_1 + \frac{\frac{3}{2}\lambda_3\tau_2 - \tau_1\frac{\beta^2}{R}}{\frac{9}{4}\lambda_3^2 - \beta^2} m_2. \tag{22}$$

Substitution into (19) yields the representation under the frame $\{m_1, m_2, m_3\}$

$$d = \lambda_3[m_3 \otimes m_3] - (R + \tau_3)m_1 \otimes m_2 - R m_2 \otimes m_1 + m_3 \otimes (\tau_1 m_1 + \tau_2 m_2). \tag{23}$$

The above representation is also unique under the similar constricts. It can be seen from the comparison that the representation under three real roots and the representation with two conjugate complex roots are different only to one sign of the term $m_1 \otimes m_2$.

## 4. Local streamline pattern (LSP) and its geometric parameters

In the representations based on the real Schur forms, an orthonormal frame must be set up first to reveal the parameter structures of velocity gradient. The frame varies point by point according to the velocity gradient, such a representation cannot be viewed as a sum decomposition because the frame is integrated. In this paper, the local streamline pattern will be used to uncover the meanings of parameters in the representations.

### 4.1 Planar flows

In planar flows, the traceless tensor of velocity gradient has standard matrix expression $\begin{pmatrix} 0 & \pm\frac{\beta^2}{R} \\ -R & 0 \end{pmatrix}$ under some orthogonal frame, with $\beta^2 = R(R + \tau)$. Following a moving point to investigate flow pattern around it, we can use the linearized local velocity field through the local velocity gradient. Under the local coordinate system defined by the chosen frame, the local streamline passing through the neighboring point $(x_0, y_0)$ satisfies the equation

$$\frac{dx(t)}{dt} = u = -Ry, \qquad \frac{dy(t)}{dt} = v = \pm\frac{\beta^2}{R}x; \quad x(0) = x_0, y(0) = y_0, \tag{24.1}$$

and can be solved as

$$x(t) = \begin{pmatrix} x_0 \cos\beta t - \frac{R}{\beta} y_0 \sin\beta t \\ \frac{1}{2}\left(x_0 + \frac{R}{\beta} y_0\right)e^{\beta t} + \frac{1}{2}\left(x_0 - \frac{R}{\beta} y_0\right)e^{-\beta t} \end{pmatrix}, \quad y(t) = \begin{pmatrix} \frac{\beta}{R} x_0 \sin\beta t + y_0 \cos\beta t \\ \frac{1}{2}\left(\frac{\beta}{R} x_0 + y_0\right)e^{\beta t} - \frac{1}{2}\left(\frac{\beta}{R} x_0 - y_0\right)e^{-\beta t} \end{pmatrix}. \tag{24.2}$$

Eliminating the parameter $t$, we obtain the local streamline pattern as:

$$x^2 \pm \frac{R}{R + \tau} y^2 = x_0^2 \pm \frac{R}{R + \tau} y_0^2, \tag{25}$$

which means that (1) for the case of two conjugate complex roots, the streamline is an ellipse centered at the origin with the semi-axis ratio $\sqrt{R/(R + \tau)}$, as shown in Fig. 2a; (2) for the case of two real roots, the streamline is a hyperbola as shown in Fig. 2b, where $\sqrt{R/(R + \tau)}$ gives the slope of characteristic line. The close ellipse of streamline is the symbol of vortex, while $\beta$ indicates angular frequency and $\tau/R$ the shape.

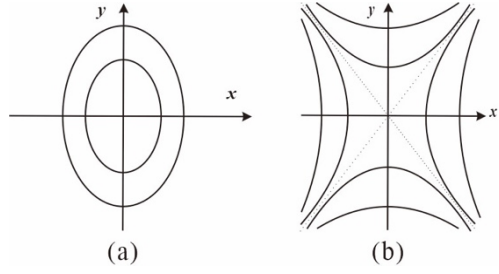

**Fig. 2.** Local streamline pattern of two-dimensional flows at point where the velocity gradient tensor has: (a) complex eigenvalues, (b) real eigenvalues.

4.2 Three-dimensional flows

In three-dimensional flows, the local streamline $x(t)$ with starting point $x(0) = x_0$ is determined by $\dot{x}(t) = x(t) \cdot d$. Zhou *et al.* (1999) figured out the streamline pattern under the affine frame $\{v_r, v_{cr}, v_{ci}\}$, and proposed the parameter $\lambda_i$ as the strength of the vortex. Now, we present the local streamline patterns under the orthonormal frames $\{m_1, m_2, m_3\}$ and $\{n_1, n_2, n_3\}$, respectively.

Under the frame $\{m_1, m_2, m_3\}$, the radius vector can be denoted by $r = x_i m_i$, and the streamline equations are

$$d = \lambda_3 \lfloor m_3 \otimes m_3 \rfloor + (R + \tau_3) m_1 \otimes m_2 - R m_2 \otimes m_1 + m_3 \otimes (\tau_1 m_1 + \tau_2 m_2)$$

$$\frac{dx_1(t)}{dt} = -\frac{\lambda_3}{2} x_1 - R x_2 + \tau_1 x_3, \quad \frac{dx_2(t)}{dt} = \pm(R + \tau_3) x_1 - \frac{\lambda_3}{2} x_2 + \tau_2 x_3, \quad \frac{dx_3(t)}{dt} = \lambda_3 x_3. \tag{26}$$

Denoted by $c_1 = \frac{\frac{3}{2}\lambda_3 \tau_1 - R \tau_2}{\frac{9}{4}\lambda_3^2 \pm \beta^2}$, $c_2 = \frac{\frac{3}{2}\lambda_3 \tau_2 \pm \frac{\beta^2}{R}\tau_1}{\frac{9}{4}\lambda_3^2 \pm \beta^2}$, these equations can be worked out

$$x_1(t) = e^{-\frac{1}{2}\lambda_3 t} \begin{pmatrix} (x_1^0 - c_1 x_3^0)\cos\beta t - \frac{R}{\beta}(x_2^0 - c_2 x_3^0)\sin\beta t \\ \frac{1}{2}\left[x_1^0 - c_1 x_3^0 - \frac{R}{\beta}(x_2^0 - c_2 x_3^0)\right]e^{\beta t} + \frac{1}{2}\left[x_1^0 - c_1 x_3^0 + \frac{R}{\beta}(x_2^0 - c_2 x_3^0)\right]e^{-\beta t} \end{pmatrix} + c_1 x_3^0 e^{\lambda_3 t}, \tag{27.1}$$

$$x_2(t) = e^{-\frac{1}{2}\lambda_3 t} \begin{pmatrix} \frac{\beta}{R}(x_1^0 - c_1 x_3^0)\sin\beta t + (x_2^0 - c_2 x_3^0)\cos\beta t \\ -\frac{1}{2}\left[\frac{\beta}{R}(x_1^0 - c_1 x_3^0) - x_2^0 + c_2 x_3^0\right]e^{\beta t} + \frac{1}{2}\left[\frac{\beta}{R}(x_1^0 - c_1 x_3^0) + x_2^0 - c_2 x_3^0\right]e^{-\beta t} \end{pmatrix} + c_2 x_3^0 e^{\lambda_3 t}, \tag{27.2}$$

$$x_3(t) = x_3^0 e^{\lambda_3 t}, \tag{27.3}$$

which means under the frame $\{m_1, m_2, m_3\}$ the streamline looks like an elliptical helix on the contracted conical surface with eccentric extension:

$$(x_1 - c_1 x_3)^2 \pm \frac{R}{R + \tau_3}(x_2 - c_2 x_3)^2 = \frac{x_3^0}{x_3}\left[(x_1^0 - c_1 x_3^0)^2 \pm \frac{R}{R + \tau_3}(x_2^0 - c_2 x_3^0)^2\right], \tag{28}$$

as shown in Fig. 3(b), where the matrix expression of velocity gradient under the frame $\{m_1, m_2, m_3\}$ is

$$D_R = \begin{bmatrix} -0.04 & 2 & 0 \\ -1 & -0.04 & 0 \\ 1 & 1 & 0.08 \end{bmatrix}.$$

In comparison with (10), the extension direction is actually along with the left eigenvector $N^3$.

If the frame changes to be $\{n_1, n_2, n_3\}$, we obtain the streamline equations

$$\frac{dx(t)}{dt} = -\frac{\lambda_3}{2} x - R^* y, \quad \frac{dy(t)}{dt} = \pm(R^* + \tau_3^*) x - \frac{\lambda_3}{2} y, \quad \frac{dz(t)}{dt} = \tau_1^* x + \tau_2^* y + \lambda_3 z \tag{29}$$

with the starting point $x(0) = x_0, y(0) = y_0, z(0) = z_0$. The explicit solutions of the streamline are:

$$x(t) = e^{-\frac{1}{2}\lambda_3 t}\begin{pmatrix} x_0 \cos\beta t - \frac{R^*}{\beta} y_0 \sin\beta t \\ \frac{1}{2}\left(x_0 - \frac{R^*}{\beta} y_0\right)e^{\beta t} + \frac{1}{2}\left(x_0 + \frac{R^*}{\beta} y_0\right)e^{-\beta t} \end{pmatrix}, \tag{30.1}$$

$$y(t) = e^{-\frac{1}{2}\lambda_3 t}\begin{pmatrix} \frac{\beta}{R^*}x_0 \sin\beta t + x_2^0 \cos\beta t \\ -\frac{1}{2}\left(\frac{\beta}{R^*}x_0 - y_0\right)e^{\beta t} + \frac{1}{2}\left(\frac{\beta}{R^*}x_0 + y_0\right)e^{-\beta t} \end{pmatrix}, \tag{30.2}$$

$$z(t) = C_3 e^{\lambda_3 t} + \frac{\tau_1^*}{2}e^{-\frac{1}{2}\lambda_3 t}\begin{pmatrix} \frac{x_0 + \iota\frac{R^*}{\beta}y_0}{-\frac{3}{2}\lambda_3 + \iota\beta}e^{\iota\beta t} + \frac{x_0 - \iota\frac{R^*}{\beta}y_0}{-\frac{3}{2}\lambda_3 - \iota\beta}e^{-\iota\beta t} \\ \frac{x_0 - \frac{R^*}{\beta}y_0}{-\frac{3}{2}\lambda_3 + \beta}e^{\beta t} + \frac{x_0 + \frac{R^*}{\beta}y_0}{-\frac{3}{2}\lambda_3 - \beta}e^{-\beta t} \end{pmatrix} - \frac{\beta\tau_2^*}{2R^*}e^{-\frac{1}{2}\lambda_3 t}\begin{pmatrix} \frac{\iota x_0 - \frac{R^*}{\beta}y_0}{-\frac{3}{2}\lambda_3 + \iota\beta}e^{\iota\beta t} - \frac{\iota x_0 + \frac{R^*}{\beta}y_0}{-\frac{3}{2}\lambda_3 - \iota\beta}e^{-\iota\beta t} \\ \frac{x_0 - \frac{R^*}{\beta}y_0}{-\frac{3}{2}\lambda_3 + \beta}e^{\beta t} - \frac{x_0 + \frac{R^*}{\beta}y_0}{-\frac{3}{2}\lambda_3 - \beta}e^{-\beta t} \end{pmatrix}, \tag{30.3}$$

with

$$C_3 = z_0 - \frac{\tau_1^*}{2}\begin{pmatrix} \frac{x_0 + \iota\frac{R^*}{\beta}y_0}{-\frac{3}{2}\lambda_3 + \iota\beta} + \frac{x_0 - \iota\frac{R^*}{\beta}y_0}{-\frac{3}{2}\lambda_3 - \iota\beta} \\ \frac{x_0 + \frac{R^*}{\beta}y_0}{-\frac{3}{2}\lambda_3 + \beta} + \frac{x_0 - \frac{R^*}{\beta}y_0}{-\frac{3}{2}\lambda_3 - \beta} \end{pmatrix} - \frac{\beta\tau_2^*}{2R^*}\begin{pmatrix} \frac{\iota x_0 - \frac{R^*}{\beta}y_0}{-\frac{3}{2}\lambda_3 + \iota\beta} - \frac{\iota x_0 + \frac{R^*}{\beta}y_0}{-\frac{3}{2}\lambda_3 - \iota\beta} \\ \frac{x_0 - \frac{R^*}{\beta}y_0}{-\frac{3}{2}\lambda_3 + \beta} - \frac{x_0 + \frac{R^*}{\beta}y_0}{-\frac{3}{2}\lambda_3 - \beta} \end{pmatrix}. \tag{30.4}$$

From the first two solutions of $x(t)$ and $y(t)$, we get the pattern of the local streamline as:

$$x^2 \pm \frac{R^*}{R^* + \tau_3^*}y^2 = e^{-\lambda_3 t}\left(x_0^2 \pm \frac{R^*}{R^* + \tau_3^*}y_0^2\right), \tag{31}$$

which means under the frame $\{\boldsymbol{n}_1, \boldsymbol{n}_2, \boldsymbol{n}_3\}$ the streamline looks like an elliptical helix on the contracted conical surface with z-axis as its center, with the same pattern as (28), say in Fig. 3(b), but the z-axis is along the extension direction and so with no explicit description of the base plane.

Now, we have three kinds of views of local streamline pattern for the same velocity gradient with only one real eigenvalue: (i) Zhou's view from the deformed frame, (ii) Li's view from the orthonormal frame of the left eigen-representation, (iii) our view from the orthonormal frame of the right eigen-representation. Under the affine frame of eigenvectors, the streamline formula (3) results in a pattern of circular helix becoming small with the parameter *t* if the real eigenvalue $\lambda_r$ is positive, the pattern equation is

$$y_1^2(t) + y_2^2(t) = e^{-\lambda_r t}[(x_1^0)^2 + (x_2^0)^2]. \tag{32}$$

The streamline formula (30) gives a pattern of elliptical helix under the frame $\{\boldsymbol{n}_1, \boldsymbol{n}_2, \boldsymbol{n}_3\}$ becoming small with the parameter *t* if the real eigenvalue $\lambda_3$ is positive, while the streamline formula (27) under the frame $\{\boldsymbol{m}_1, \boldsymbol{m}_2, \boldsymbol{m}_3\}$ presents a pattern of elliptical helix with monotonous z-position and eccentrically contracted with the z-coordinate.

It is noticeable that the LSP is objective, namely the streamline passing through the same point is unique whatever the frame and the corresponding representation are chosen. The question is in which representation the derived parameters makes up a good geometric description of the local streamline pattern? After careful investigation, we find the following geometric features of LSP:

(1) There is a plane through the origin, called the base plane with $\boldsymbol{m}_3$ as its normal, except the streamlines on the plane no streamline goes through this plane. The unit vector can be made unique by the sign of $\beta$ in (7) or $R$ in (8).

(2) There are two typical directions for the vortical streamlines: one is the rotation axis defined by $\boldsymbol{m}_3$, another is the extension direction defined by $\boldsymbol{N}^3$ adjusted through $c_1, c_2$ or $\tau_1, \tau_2$ in the frame $\{\boldsymbol{m}_1, \boldsymbol{m}_2, \boldsymbol{m}_3\}$ as in (10) or directly $\boldsymbol{n}_3$ along which the streamline is stretch far away the base plane ($\lambda_3 > 0$) or compressed close to the base plane ($\lambda_3 < 0$).

(3) Any streamline out of the base plane lies on an elliptical cone with the extension direction as its axis, spirals up ($\lambda_3 > 0$) or down ($\lambda_3 < 0$) along the rotation axis $\boldsymbol{m}_3$. The speed of rotation is determined by the parameter $\beta$, as pointed out by Zhou *et al* (1999).

(4) The streamline on the base plane rotates to or away from the center, namely the origin, in the form of ellipse with $\tau_3/R$ indicating the ellipticity. Other streamlines projected onto the base plane have the same the form of ellipse, but eccentricity occurs linearly according to the distance from the base plane, and the size is inversely proportional to the height from the base plane.

Therefore, besides the basic parameters $\lambda_3, \beta$, it is very clear that the parameter set $\{R, \tau_1, \tau_2, \tau_3\}$ of the right eigen-representation perfectly indicates the geometric features of LSP. As a contrast, the parameter set $\{R^*, \tau_1^*, \tau_2^*, \tau_3^*\}$ of the left eigen-representation has no such obvious geometric meaning, except that $\tau_3^*/R$ indicating the ellipticity of shapes cut from the cone of the streamline with the planes perpendicular to the extension direction $\mathbf{n}_3$. The direction $\mathbf{n}_3$ Li *et al*. and Liu *et al*. refer to as the rotation axis is actually the extension direction of local streamlines, that has been pointed out by Zhou *et al*. The real rotation axis of local streamlines is $\mathbf{m}_3$, which was also realized by Zhou *et al*, but not defined explicitly. When the parameters $\tau_1, \tau_2$ vanish, the right eigen-representation becomes identical to the left eigen-representation, the difference between two eigen-representation can be characterized by the (acute) included angle between $\mathbf{N}_3$ and $\mathbf{N}^3$, the cosine of it has formula:

$$cos(\mathbf{N}_3, \mathbf{N}^3) = \frac{\frac{9}{4}\lambda_3^2 + \beta^2}{\sqrt{\left(\frac{9}{4}\lambda_3^2 + \beta^2\right)^2 + \left(\frac{3}{2}\lambda_3\tau_1 - R\tau_2\right)^2 + \left(\frac{3}{2}\lambda_3\tau_2 + \frac{\beta^2}{R}\tau_1\right)^2}}. \tag{33}$$

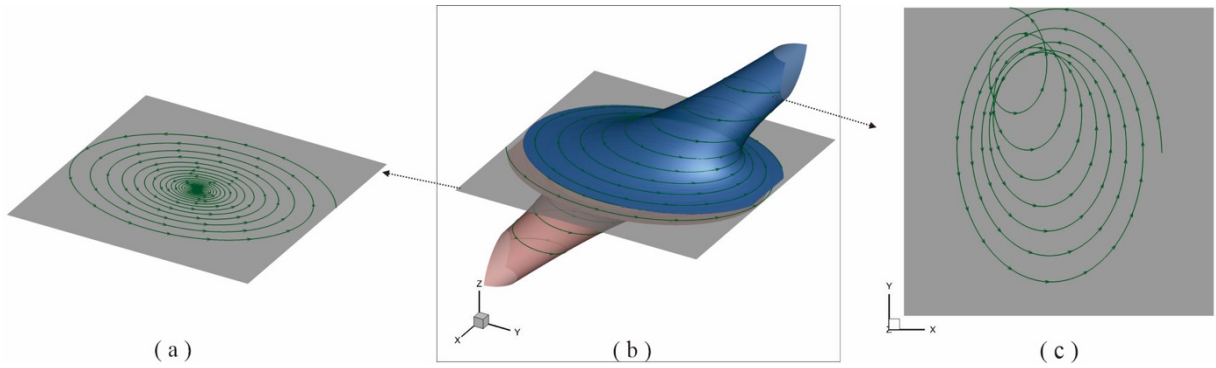

**Fig. 3**. Local streamline pattern of velocity gradient with a unique real eigenvalue: (a) the base plane and the streamline on it, (b) local streamline pattern, (c) the projection of an above streamline on the base plane.

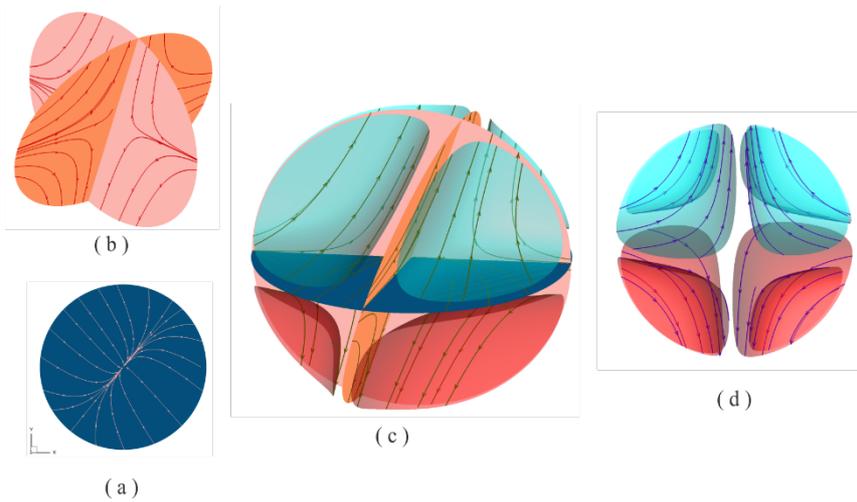

**Fig. 4**. Local streamline pattern of velocity gradient with three real eigenvalues: (a) the foci plane of streamlines normal to $\mathbf{m}_3$, namely the eigenvectors related to the eigenvalue with the maximal module, (b) two saddle planes of streamlines normal to the eigenvectors related to two eigenvalues with smaller modules, (c) local streamline pattern divided by three base planes, (d) local streamline pattern removing three base planes.

The local streamline pattern for the velocity gradient with three real eigenvalues has more isolated regions of streamlines. For the case of one eigenvalue having a maximal module, say the velocity gradient tensor has the matrix expression

$$\mathbf{D}_R = \begin{bmatrix} -0.5 & -0.3 & 0 \\ -0.2 & -0.5 & 0 \\ 0.3 & 0.3 & 1.0 \end{bmatrix}$$

under the frame $\{\boldsymbol{m}_1, \boldsymbol{m}_2, \boldsymbol{m}_3\}$, the local streamline pattern is divided into eight regions, as illustrated in Fig. 4(c). Besides the base plane characterized by $\boldsymbol{m}_3$ as the foci plane of streamlines (see Fig. 4(a)), the other two coordinate planes defined by the rest eigenvectors also keep the streamlines on the planes, but in the form of saddle plane, see Fig. 4(b). Any streamline not on the coordinate planes will be confined in eight isolated regions (Fig. 4(d)), each streamline is located on a hyperboloid, as formulated in (28).

## 5. Discussion and Case study

### 5.1 Discussion

Let's see a simple example first. The Taylor-Couette flow between two concentric cylinders is the simplest curved flow. In the case of layer flow with the boundary conditions $v_\theta(r_1) = \Omega_1 r_1, v_\theta(r_2) = \Omega_2 r_2$, the exact solution from the N-S equations reads

$$v_\theta(r) = \frac{\Omega_2 r_2^2 - \Omega_1 r_1^2}{r_2^2 - r_1^2} r + \frac{(\Omega_1 - \Omega_2) r_2^2 r_1^2}{r_2^2 - r_1^2} r^{-1}, \tag{34}$$

which yields the velocity gradient as

$$\boldsymbol{d} = \frac{\Omega_2 r_2^2 - \Omega_1 r_1^2}{r_2^2 - r_1^2} (\boldsymbol{e}_r \otimes \boldsymbol{e}_\theta - \boldsymbol{e}_\theta \otimes \boldsymbol{e}_r) - \frac{(\Omega_1 - \Omega_2) r_2^2 r_1^2}{r_2^2 - r_1^2} r^{-2} (\boldsymbol{e}_r \otimes \boldsymbol{e}_\theta + \boldsymbol{e}_\theta \otimes \boldsymbol{e}_r). \tag{35}$$

For simplicity, set $r_2 = 2r_1 = \sqrt{2} r_0$, $\Omega_2 = -\Omega_1 = \Omega_0$, the velocity gradient reduces to

$$\boldsymbol{d} = \left(\frac{4}{3}\frac{r_0^2}{r^2} + \frac{5}{3}\right) \Omega_0 \boldsymbol{e}_r \otimes \boldsymbol{e}_\theta + \left(\frac{4}{3}\frac{r_0^2}{r^2} - \frac{5}{3}\right) \Omega_0 \boldsymbol{e}_\theta \otimes \boldsymbol{e}_r. \tag{36}$$

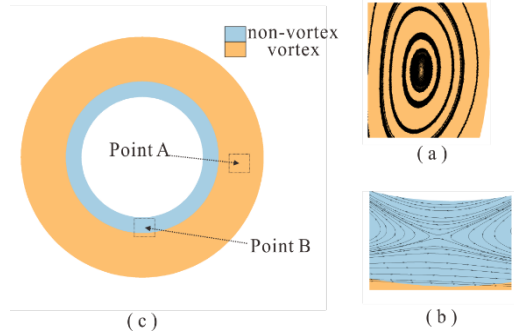

**Fig. 5**. The property of Liutex in the Taylor-Couette flow: (a) local streamlines for the velocity gradient at point A in vortex region, (b) local streamlines for the velocity gradient at point B in non-vortex region, (c) division of vortex and non-vortex regions.

According to the classical model, this flow has constant angular velocity of rotation $\frac{5}{3}\Omega_0$ everywhere, but the model of LSP shows that the fluid has no rotation when $r < 0.8944 r_0$ ($r_1 = r_0/\sqrt{2}$), and variable angular velocity elsewhere, as shown in Fig. 5. Constant rotation, varying rotation or even no rotation, which one is more intuitive?

The use of LSP in the identification of vortex emphasizes that there is no vortex in the straight flows, which can be said to be a breakthrough to the classical viewpoint of flow motion. The real Schur forms of velocity gradient starts with the eigenvector of the real eigenvalue, and brings insight into the local streamline pattern investigated by the observer moving with the local fluid particle. The precondition to associate this viewpoint with the vortex is that the vortex is carried along by the mainstream and so it is more suitable to investigate the vortex following the current particle. But such a presupposition is basically wrong, and will bring illusion. We would like to point out several fallacies in the use of LSP viewpoint:

(1) Sum decomposition: the eigen-representations are not a kind of sum decomposition, because the algebraic structures of its components indicated by parameters, say $\lambda_3, R, \tau_1, \tau_2, \tau_3$, are not invariant under frame transformation.

(2) Rigid-body rotation: there are streamlines revolving around some center, but no fluid element rotates like a rigid body, even in the average sense.

(3) Frame transformation: the orthonormal frame is derived from the velocity gradient, which varies point by point. It is

unsuitable to move the transformation inside the gradient operator, say $\boldsymbol{D}_L = \boldsymbol{P}d\boldsymbol{P}^T \neq \boldsymbol{P}\nabla(\boldsymbol{P}\boldsymbol{v}) = \boldsymbol{P}\nabla(\boldsymbol{V})$.

Now we present an example to show the invariance of local streamline. For a particle, recast at the origin with zero velocity, has the velocity gradient in the right real Schur form

$$\boldsymbol{D}_R = \begin{bmatrix} -0.0193773 & 0.240057 & 0.0 \\ -0.113056 & -0.0193773 & 0.0 \\ 0.063467 & 0.099629 & 0.0387546 \end{bmatrix}. \tag{37}$$

The basic parameters read $\lambda_3 = 0.0387546, \beta = 0.164742$, and the parameters in the right eigen-representation are $R = 0.113056, \tau_1 = 0.063467, \tau_2 = 0.099629, \tau_3 = 0.127001$. Under the frame $\{\boldsymbol{m}_1, \boldsymbol{m}_2, \boldsymbol{m}_3\}$, the streamline passing through the point $(2.97518, -4.30261, 0.1)$ is shown in Fig. 6(a). After a frame transformation

$$(\boldsymbol{n}_1, \boldsymbol{n}_2, \boldsymbol{n}_3) = (\boldsymbol{m}_1, \boldsymbol{m}_2, \boldsymbol{m}_3)\begin{pmatrix} 0.916866 & 0.345346 & -0.200229 \\ -0.183951 & 0.810662 & 0.555868 \\ 0.354286 & -0.472825 & 0.806795 \end{pmatrix}, \tag{38}$$

we obtain the left real Schur form of the velocity gradient under the frame $(\boldsymbol{n}_1, \boldsymbol{n}_2, \boldsymbol{n}_3)$

$$\boldsymbol{D}_L = \begin{bmatrix} -0.0193773 & 0.212251 & 0.149917 \\ -0.127868 & -0.0193773 & 0.0220819 \\ 0 & 0 & 0.0387546 \end{bmatrix}. \tag{39}$$

the local streamline passing through the same spatial point, namely $(3.55474, -2.50777, -2.90672)$ in the new frame, is illustrated in Fig. 6(b). It is obvious that two streamlines are actually the same (Fig. 6(c)).

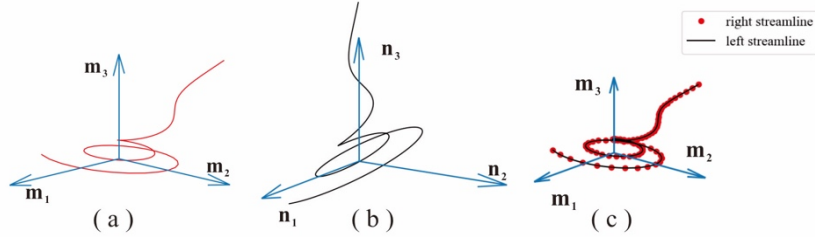

**Fig. 6**. A local streamline (a) viewed under the frame $\{\boldsymbol{m}_1, \boldsymbol{m}_2, \boldsymbol{m}_3\}$ passing through the point $(2.97518, -4.30261, 0.1)$, and (b) viewed under the frame $(\boldsymbol{n}_1, \boldsymbol{n}_2, \boldsymbol{n}_3)$ passing through the point $(3.55474, -2.50777, -2.90672)$. Two views are completely merged in (c).

### 5.2 Case study

In this subsection, we try to verify the relation between the right and left eigen-representations by analyzing the vortex features in the late transition process of the flat boundary flow with Mach number of 0.5, through the DNS data provided by Liu Chaoqun's research group, as shown in Fig. 7.

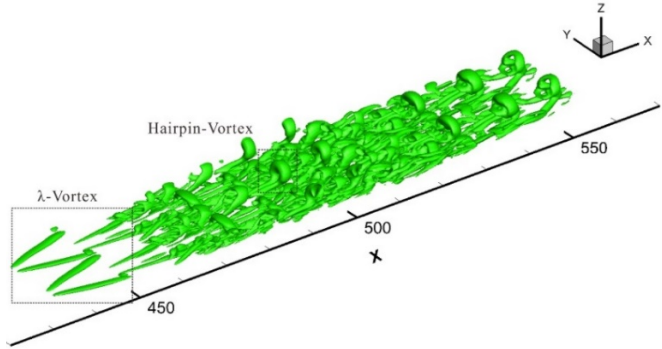

**Fig. 7.** The DNS data of the flat boundary flow with Mach number of 0.5, shown with criterion $\Omega = 0.52$.

Theodorsen (1952) pointed out that the horseshoe-shape vortex, now usually called the hairpin vortex, is the basic structure in the turbulent boundary layer flow. We pick up a typical hairpin vortex in the above DNA data, and find that the unit vectors $\boldsymbol{m}_3$ and $\boldsymbol{n}_3$ are apparently different in its leg part, while they tend to be consistent in its head part and at the junction of the vortex leg with the vortex ring, as shown in Fig. 8(left). Randomly taking 30 samples of R and R* from the leg and head, respectively, a stable positive correlation between them is obvious as predicted by the formulae (17) and (18), or shown in Fig. 8(right).

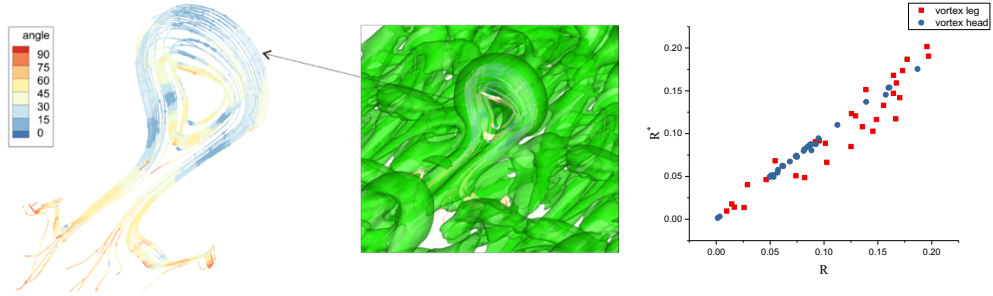

**Fig. 8**. The distribution of included angle between $m_3$ and $n_3$ on a typical hairpin-vortex. In the left Liutex lines extracted from the middle field with typical vortices, the blue color in the head and junction shows the small included angle while the red color mainly in the leg shows the large included angle. The right is the correlation between R and $R^*$ using 30 random samples from the leg and head, respectively.

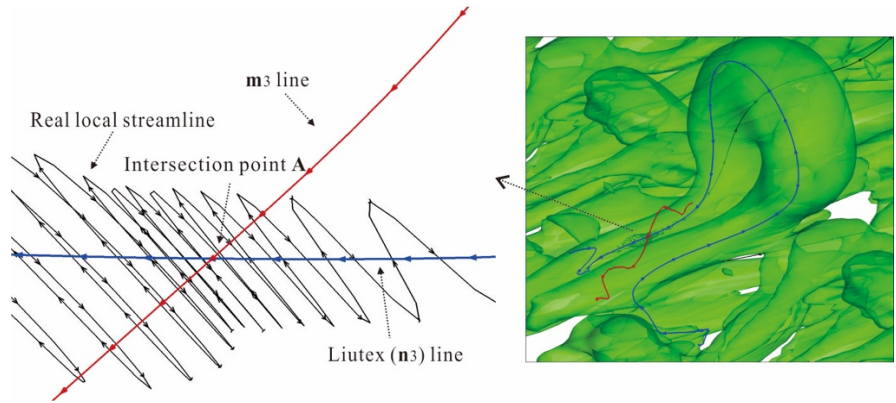

**Fig. 9**. A real local streamline and its direction features observed following a fluid particle in the leg of a typical vortex.

Finally, we take a point **A** from the vortex leg, which has coordinates $(488.114, 11.6319, 2.25222)$ in the DNS data and the velocity at this point is $v_A = (0.847403, -0.0226804, -0.0415701)$, and the velocity gradient

$$d_A = \begin{bmatrix} 0.0321266 & -0.0202696 & -0.000168884 \\ 0.169304 & 0.000339247 & -0.258384 \\ -0.146806 & 0.207205 & -0.0341335 \end{bmatrix}. \tag{40}$$

The real local streamlines from the original flow field, in comparison the LSPs discussed before are about the local linearized field, are observed by following the motion of the particle **A**. A real local streamline through the point a little deviating from the intersection point **A**, say $(488.114, 11.5319, 2.25222)$, is illustrated in Fig. 9, where the $m_3$ and $n_3$ (Liutex) lines through the point **A** are also presented. It is very clear that the local rotation axis at the point **A** is $m_3$ while the extention of the local streamline is well consistent with the Liutex line. Such an observation completely coincides with the LSP analysis in the previous section.

## 6. Conclusions

In this paper, the proposed vortex identification methods based on the real Schur forms of the velocity gradient have been revisited and developed. The tensor forms of left/right real Schur forms, called the eigen-representations now, are obtained and the relations between them are clarified, while a similar eigen-representation in the case of three real roots is also proposed. The right eigen-representation is recommended through the investigation on the geometric meaning of parameters, which is based on the analysis of the local streamline pattern (LSP). By the discussion and case study, we conclude that a presupposition needs to be further confirmed that all vortices are carried along by the current velocity.

The following viewpoints would be important for the future studies:

(1) The eigen-representation methods are non-classical and cannot be analyzed by a modified deformation model.

(2) Different eigen-representations present the same LSP, with features of a rotation direction/speed and a deviated

extension direction, and a projection shape in the base plane, while $\beta$ is more suitable to indicate the rotation speed.

(3) Since the parameters in the right eigen-representations are more sensible to the geometrical features of the LSP, the right eigen-representation is better in characterizing the vortex features of the LSP.

The original intention of making use of the eigen-representations is to reflect the intuition that there is no vortex in the straight laminar flow, but the eigen-representations destroy the spirit of classical deformation analysis. It is questionable whether a more reasonable explanation can be proposed in the vortex identification of the Taylor-Couette flow between two concentric cylinders. However, in any case, the analysis of LSP based on velocity gradient breaks through the classical deformation analysis and strives to pursue an authenticity description of the vortex to some extent, which is worthy of affirmation. In fact, there is a more direct and novel thinking, that is, "the rotation of fluid happens in any curved flow, but the vortex begins with a core of rotation", which will be discussed in another paper of ours.

## Appendix A

Making use of notation (17.2), the eigenvectors $N_1, N_2$ in (11) can be written in terms of $m_1, m_2, m_3$ by

$$N_1 = \sqrt{\frac{1}{2}}(c_0 m_1 - c_0^{-1} m_2 - (c_0 c_1 - c_0^{-1} c_2) m_3), \qquad N_2 = \sqrt{\frac{1}{2}}(c_0 m_1 + c_0^{-1} m_2 - (c_0 c_1 + c_0^{-1} c_2) m_3), \qquad (A.1)$$

and $N_3 = m_3$. We will deduce the orthonormal bases $n_1, n_2, n_3$ and parameters $R^*, \tau_1^*, \tau_2^*, \tau_3^*$ in the left eigen-representation. In virtue of the notations

$$A^2 = N_1 \cdot N_1 = \frac{1}{2}[c_0^2(1 + c_1^2) + c_0^{-2}(1 + c_2^2) - 2c_1 c_2], \qquad (A.2.1)$$

$$B^2 = N_2 \cdot N_2 = \frac{1}{2}[c_0^2(1 + c_1^2) + c_0^{-2}(1 + c_2^2) + 2c_1 c_2], \qquad (A.2.2)$$

$$N_1 \cdot N_2 = AB\cos\gamma = \frac{1}{2}[c_0^2(1 + c_1^2) - c_0^{-2}(1 + c_2^2)]. \qquad (A.2.3)$$

we first isomorphize the modules of the real part and imaginary part of $N^1 \pm \iota N^2$ by a planar rotation, namely two vectors of the real part and imaginary part of

$$(N_1 + \iota N_2)e^{-\iota\theta} = (N_1 \cos\theta + N_2 \sin\theta) + \iota(N_2 \cos\theta - N_1 \sin\theta) \qquad (A.3)$$

have the same size, yielding the unique rotation angle $\theta$ by

$$\tan 2\theta = \frac{B^2 - A^2}{2AB\cos\gamma} = \frac{2c_1 c_2}{c_0^2(1 + c_1^2) - c_0^{-2}(1 + c_2^2)}, \qquad -\frac{\pi}{2} < \theta \leq \frac{\pi}{2} \qquad (A.4)$$

if the included angle between $n_1$ and $m_1$ is confined to be acute, namely $m_1 \cdot n_1 > 0$. Then using the formula

$$n_{1,2} = \frac{N_2 \pm N_1}{|N_2 \pm N_1|} \sim \frac{N_2 \cos\theta - N_1 \sin\theta \pm (N_1 \cos\theta + N_2 \sin\theta)}{|N_2 \cos\theta - N_1 \sin\theta \pm (N_1 \cos\theta + N_2 \sin\theta)|}, \qquad (A.5)$$

we obtain

$$n_1 = \frac{\sqrt{2}[c_0 m_1 \cos\theta + c_0^{-1} m_2 \sin\theta - (c_0 c_1 \cos\theta + c_0^{-1} c_2 \sin\theta) m_3]}{\sqrt{A^2 + B^2 + \sqrt{(B^2 + A^2)^2 - 4A^2 B^2 \sin^2\gamma}}}, \qquad (A.6.1)$$

$$n_2 = \frac{\sqrt{2}[-c_0 m_1 \sin\theta + c_0^{-1} m_2 \cos\theta + (c_0 c_1 \sin\theta + c_0^{-1} c_2 \cos\theta) m_3]}{\sqrt{A^2 + B^2 - \sqrt{(B^2 + A^2)^2 - 4A^2 B^2 \sin^2\gamma}}}, \qquad (A.6.2)$$

and consequently

$$n_3 = n_1 \times n_2 = \frac{2(m_3 + c_1 m_1 + c_2 m_2)}{A^2 + B^2 + \sqrt{(B^2 + A^2)^2 - 4A^2 B^2 \sin^2\gamma}} = \frac{m_3 + c_1 m_1 + c_2 m_2}{\sqrt{1 + c_1^2 + c_2^2}}, \qquad (A.6.3)$$

which coincides with (10). According to the definition (12), we can calculate

$$\tau_1^* = n_1 \cdot d \cdot n_3, \qquad \tau_2^* = n_2 \cdot d \cdot n_3, \qquad R^* = -n_2 \cdot d \cdot n_1, \qquad \tau_3^* = \frac{\beta^2}{R^*} - R^*. \qquad (A.7)$$

There is a more direct formula of $R^*$ from the expressions (13), say

$$R^* = \beta \frac{|N_1 cos\theta + N_2 sin\theta - N_2 cos\theta + N_1 sin\theta|}{|N_1 cos\theta + N_2 sin\theta + N_2 cos\theta - N_1 sin\theta|}. \quad (A.8)$$

Substitution of (A.2) and (A.4) yield

$$R^{*2} = \beta^2 \frac{A^2 + B^2 - \sqrt{(B^2 + A^2)^2 - 4A^2 B^2 \sin^2 \gamma}}{A^2 + B^2 + \sqrt{(B^2 + A^2)^2 - 4A^2 B^2 \sin^2 \gamma}}, \quad (A.9)$$

or expressed by as (17) in terms of the parameters $R, \tau_3, c_0, c_1, c_2$.